# Ein Typenrad auf der Überholspur

## Die Kult-Schreibmaschine „Erika" trifft KI

*Karola Köpferl, Albrecht Kurze*

Technische Universität Chemnitz, Deutschland
{karola.koepferl, albrecht.kurze}@informatik.tu-chemnitz.de

**Abstract**

Im 15. Jahrhundert revolutionierte der Buchdruck die Verbreitung von Informationen. Innovationen wie Schreibmaschinen und Computer haben im Laufe der Zeit die Geschwindigkeit und das Volumen der Informationsströme erhöht. Neuere Entwicklungen großer Sprachmodellen wie ChatGPT ermöglichen die Generierung von Text in Sekundenschnelle. Vielen Menschen ist jedoch nicht klar, wie das funktioniert und welche Auswirkungen das langfristig hat. Daher haben wir eine alte Schreibmaschine „gehackt", sodass die Nutzer:innen darüber mit einem LLM-Chatbot interagieren können, was inzwischen über 1200 Teilnehmer:innen erleben konnten. Es hilft, die Möglichkeiten und Grenzen von KI zu verstehen. Es gibt uns Forschenden Einblicke in die Konzepte der Teilnehmer:innen über KI sowie ihre Erwartungen und Bedenken. Es wirft Fragen zu diesen technologischen Entwicklungen auf und regt Diskussionen über die gesellschaftlichen Auswirkungen der Verdichtung und Beschleunigung von Informations- und Kommunikationsströmen an.

**Keywords:** Technostalgia; Chatbots; KI; LLM; Kommunikationskultur; Human-Computer Interaction

## 1 Einleitung

Der globale Informations- und Wissenstransfer über digitale Datennetze, ermöglicht durch Computer und das Internet, ist ein grundlegender Aspekt heutiger Informationsgesellschaft (Seefried 2024). Diese Entwicklung baut auf der Geschichte von Kommunikationskulturen einerseits und Fortschritten der Informationstechnik andererseits auf (Technische Sammlungen Dresden 2000). Die Kulturtechnik des Schreibens und der Textproduktion hat sich im Laufe der Jahrhunderte entwickelt und ist zu einem festen Bestandteil des menschlichen Kulturerbes geworden.

Vom historischen Pult im Kloster über das Großraumbüro bis zum digitalen Nomaden in der Hängematte: die Fähigkeit, Ideen, Erfahrungen und Informationen durch Sprache und Schrift zu vermitteln, ist eine wesentliche Grundlage der Entwicklung und des Erhalts menschlicher Gesellschaften. Immer wieder tauchen Aspekte der Verdichtung von Text auf – immer mit einer Form der Beschleunigung der Arbeit in der Informationsverbreitung, (fast) immer mit der physischen Reproduktion auf dem Medium Papier.

Die Erfindung des Buchdrucks durch Gutenberg im 15. Jahrhundert ermöglichte die Massenproduktion von Büchern und revolutionierte die Verbreitung von Informationen (Lehmann-Haupt 2024). Die Einführung der Schreibmaschine gegen Ende des 19. Jahrhunderts führte zu einer starken Zunahme des Papierverbrauchs und der Informationsdichte und markierte einen weiteren Wendepunkt in der Geschichte der Kommunikationskultur (Technische Sammlungen Dresden 2000). Die Entwicklung von Computern und Drucktechnologien im 20. Jahrhundert war ein weiterer Meilenstein in der Kommunikationsgeschichte und führte zu einer erheblichen Zunahme des Informationsflusses (Bösch 2018). Die Begriffe Informationsfluss und Informationsstrom werden synonym verwendet und beschreiben den Weg einer Information von einem Sender zu einem oder mehreren Empfängern – in der Informationstechnik spricht man von Datenfluss.

Die Idee des KI-Chatbots stammt aus den 1950er-Jahren (Eliza Archaeology Project 2024). Doch erst die Veröffentlichung von ChatGPT durch OpenAI im November 2022 (Heaven 2023; Perlman 2022) mit der „Fähigkeit", wie von Geisterhand teilweise ausgearbeitete Texte zu generieren, hat generative KI ins öffentliche Bewusstsein gerückt.

ChatGPT und Co. haben die Art und Weise, wie wir mit Informationen interagieren und Kommunikation gestalten, erneut verändert. Täglich werden Textmengen produziert, die ausgedruckt bis zum Mond reichen würden. Wir beobachten eine Zunahme der Informationsmenge pro Zeiteinheit und eine Abnahme der benötigten Zeit pro Informationsmenge (Rosa 2005). Diese Entwicklung scheint kaum fassbar.

Sie wirft Fragen nach den sozialen und kulturellen Auswirkungen dieser Informationsmengen auf:
- Welche Perspektiven und Orientierungen haben Menschen bei der Nutzung von KI und LLMs wie ChatGPT?

- Welche gesellschaftlichen Implikationen ergeben sich aus der Verdichtung und Beschleunigung von Informationsströmen?

## 2    Methodik

Um diese Fragen zu untersuchen, wurde eine 40 Jahre alte elektronische Schreibmaschine „Erika" „gehackt" (Wahl 2023). Als Schnittstelle für die getippte Ein- und Ausgabe auf Papier zu einer LLM-API, die eine Konversation mit einem LLM-Chatbot ermöglicht und als Gesprächseinstieg über alte und neue Informations- und Kommunikationstechnologien, Schreibmaschinen und KI dient. Die elektronische Schreibmaschine „Erika S 3004" wurde Mitte der 1980er-Jahre in der DDR produziert. Diese Schreibmaschinen waren für ihre Qualität bekannt. Ihre für die damalige Zeit fortschrittlichen Funktionen machten sie weit verbreitet und beliebt. Mit über acht Millionen verkauften Exemplaren wurde „Erika" zur meistverkauften Schreibmaschinenmarke in Deutschland.

Die Schreibmaschine wurde folgendermaßen in eine „KI-Schreibmaschine" umgewandelt: Die serielle Schnittstelle der Schreibmaschine wurde Ende der 1980er-Jahre für den Betrieb als Drucker verwendet (Pohlers 2005). Daran wird ein ESP32 Mikrocontroller mit sechs Kabeln angeschlossen, einschließlich der seriellen Kommunikation, der Flusssteuerung und der Stromversorgung des Mikrocontrollers. Dieser wird per WLAN mit dem Internet verbunden, verwendet die auf der Schreibmaschine getippten Zeichen für die Eingabe auf Papier und parallel als Eingabe für einen Prompt, schickt den Prompt als Anfrage an ein LLM (Open AI ChatGPT-3.5-turbo/ChatGPT4.0-turbo Mixtral 8x7b und Llama 2.5/3), empfängt die Antwort und schickt sie zum Ausdrucken an die Schreibmaschine. Während heutiger Text in Unicode vorliegt, verwendet die Schreibmaschine einen alten 8-Bit-Zeichen- und -Befehlssatz.

## 3    Ergebnisse

Wir präsentieren die „KI-Schreibmaschine" regelmäßig auf öffentlichen Veranstaltungen und haben inzwischen Daten (Beobachtungen, Interviews und die getippten Prompts und Antworten selbst) von fast 1.200 Nutzungen durch Teilnehmende aller Altersgruppen gesammelt.

Wir beobachten folgende Effekte:

- **KI ohne Monitor:** Menschen können ohne moderne Ein- und Ausgabegeräte mit einem Chatbot kommunizieren, der ESP32-Zusatz ist klein, was die Nutzer:innen zunächst gar nicht bemerken. Mensch und Maschine tippen auf Papier: KI wird greifbar, fühlbar und die Menge/Länge der Ergebnisse wird haptisch und sichtbar.
- **Technostalgie und technische Zukunft:** Die Schreibmaschine klappert „wie früher". Damit erregt sie in Ausstellungen Aufsehen und eröffnet Dialoge über Technik in der Vergangenheit, über den Umgang mit „moderner" Technik heute und über Zukunftsfragen: Welche Berufe „lohnen" noch, wenn Textarbeit scheinbar überflüssig wird? Braucht es KI, um noch mehr Text zu produzieren?
- **Vertrauen und Wahrhaftigkeit in der Kommunikation:** Es stellt sich die Frage nach Vertrauen. Welcher Text ist noch „echt", noch „menschlich"? Nutzende fragen, ob man sicher sein kann, dass die Informationen korrekt und vertrauenswürdig sind?

Folgende Kategorien analysieren wir aus den Prompts der Nutzenden:

- **KI in der Maschine hinterfragen:** Teilnehmer:innen stellen Fragen zur Funktionsweise und den Eigenschaften des Chatbots. Beispiele: „Wer bist du?" oder „Wie hoch ist dein IQ?"
- **Nutzung wie eine Suchmaschine:** Teilnehmer:innen nutzten den Chatbot, um einfache Fragen zu stellen, ähnlich wie bei einer Suchmaschine.
- **Beratung:** Teilnehmer:innen suchten Rat zu alltäglichen, persönlichen und gesundheitlichen Fragen.
- **Kreative Aufgaben:** Chatbot soll kreative Inhalte wie Gedichte oder Briefe generieren.
- **Zukunft und Prognosen:** Teilnehmer:innen erwarten, dass die KI Zukunftsprognosen erstellen kann: „Wie wird das Wetter morgen?"
- **Provokative Prompts:** Teilnehmende testeten die Grenzen der KI mit provokativen oder illegalen Fragen – „… wie baue ich eine Bombe?"

## 4 Reflexion und Ausblick

Das Projekt "Erika" ermöglicht einen praktischen Zugang und eröffnet damit einen Dialog über den KI-Hype. Es regt den Dialog über die Entwicklung der Technologie und ihre Auswirkungen auf unser Leben an und eröffnet damit Diskussionen darüber, wie wir mit der zunehmenden Geschwindigkeit und Menge an Informationen umgehen und welche Rolle KI in unserem Alltag spielen sollte. Die Reaktionen einiger Nutzer:innen auf die KI-Schreibmaschine zeigen, dass es ein großes Interesse an den ethischen und praktischen Implikationen der KI-Nutzung gibt. Dies betrifft Fragen der Datensicherheit, der Transparenz von KI-Algorithmen und der Verantwortung für die Entwicklung und Nutzung. Der Einsatz nostalgischer Technik eröffnet zudem eine zusätzliche Dimension: Jung und Alt in diesen Fragen und Aspekten zu verbinden. Im weiteren Projektverlauf soll die Frage nach den gesellschaftlichen Implikationen der Kommunikationskultur mit Rosas gesellschaftstheoretischen Arbeiten zur Beschleunigung in der „Spätmoderne" (Rosa 2005) verknüpft werden.

## Literatur